# Challenges of identifying putative planetary-origin meteorites composed of non-igneous material

Yana Anfinogenova,[1]* John Anfinogenov[2]

**Affiliations:**

[1]Yana Anfinogenova, Ph.D., National Research Tomsk Polytechnic University (Address: 30 Lenin Ave., Tomsk 634050, Russia)

[2]John Anfinogenov, Tunguska Nature Reserve, Ministry of Natural Resources and Ecology of the Russian Federation (Address: 8 Moskovskaya Str., Vanavara, Evenki District, Karsnoyarsk Kray 648490, Russia)

*Corresponding author: Yana Anfinogenova. Address: TPU, 30 Lenin Ave., Tomsk, 634050, Russia. Tel: +79095390220. E-mail: anfiyj@gmail.com and anfy@tpu.ru

**Disclosure.** Authors declare that they do not have any financial conflict of interests associated with this article.

**Abstract.** This concept article discusses the challenges of identifying planetary-origin meteorites of non-igneous composition, primarily of sedimentary origin, distinct from SNC meteorites. The paper reviews evidence on putative sedimentary-origin meteorites and potential parent bodies for sedimentary meteorites. Authors conclude that the list of candidate parent bodies for sedimentary meteorites includes, but is not limited by the Earth, Mars, Enceladus, Ganymede, Europa, and hypothetical planets that could exist between orbits of Mars and Jupiter in the past. Authors argue that extraterrestrial origin and a parent body for meteoritic sedimentary rocks may be identified based on the entire body of evidence which is not limited solely by tests of oxygen and noble gas isotopes whose signatures may undergo terrestrial contamination and may exhibit significant heterogeneity within the Solar system and within the parent cosmic bodies. Observed fall of a cosmic body, evidence of hypervelocity fall, signs of impact in target, and the presence of fusion crust, melting, and/or shock deformation features in the fragments should be considered as priority signs of meteoritic origin.

**Highlights**

The macroscopic candidate for sedimentary meteorite is reported

Parent bodies for putative sedimentary meteorites are discussed

Parent bodies include Earth, Mars, Enceladus, Ganymede, Europa, and other objects

**Introduction**

The SNC meteorites are a group of petrologically similar achondrites that comprise the shergottites, nakhlites, and chassignites named after the locations where they were originally found: Shergotty (India), Nakhla (Egypt), and Chassigny (France) [Sohl and Spogn 2011]. All SNC meteorites display igneous features [Papike *et al.* 2009]. Till now, SNC group has been the only commonly accepted group of planetary-origin meteorites most probably coming from Mars.

Experts in sedimentary geology Ashley G. M. and Delaney J. S. [1999], however, believe that sedimentary meteorites should be also sampled on Earth in proportion to the fraction of Mars covered by sedimentary rocks. They present compositions of Barnacle Bill and Yogi from Sagan Station, Mars on classification diagram for igneous rocks with composition ranges of typical sedimentary rocks superimposed. According to the diagram, SNC meteorites, the only rocks generally accepted to form on the planet Mars, represent only a small portion of what types of meteorites from this planet might be found [Ashley *et al.* 1999]. Ashley G. M. and Delaney J. S. [1999] emphasize that fusion crust is crucial for recognition of extraterrestrial origin of meteorites and state that "if a consolidated siliciclastic sediment were ejected from Mars, the fusion crust formed during its deceleration and descent to Earth could be quite unlike anything that previous meteoritic experience defines as true fusion crust" [Ashley *et al.* 1999].

In the present paper, we discuss challenges of identifying putative sedimentary meteorites and review potential parent bodies for such rocks; isotopic heterogeneity of unmixed silicate reservoirs on Mars; possible terrestrial loss or contamination in the noble gas signatures in meteorites that spent time in the extreme weather conditions; and cosmogenic isotopes and shielding hampering identification of new type meteorites. We discuss possible nature of the first macroscopic candidate meteorite of a new type (meteorite composed of planetary sediments) from Tunguska and emphasize the significance of studying this phenomenon.

Sedimentary meteorites, found on Earth, may significantly contribute to elucidating the history of the Solar System and to the search for possible extraterrestrial life forms. Recently, a new discipline of astrobiology emerged to address fundamental questions about life in the universe. Along with developing capabilities in biosciences, informatics, and space exploration [Morrison 2001], discovery, identification, and thorough examination of putative sedimentary meteorites may help to address the astrobiology questions and to facilitate further exploration of the Solar System.

**Simulation modeling experiments**

In early 2000s, an international team performed simulation modeling experiments aimed at identifying the effects of thermal alteration during atmospheric entry of Martian analogue sediments using a robotic spacecraft of Foton series employed by Russia and the European Space Agency [Foucher et al. 2009; Brack et al. 2002]. In particular, a 3.5-byr-old sediment with present microfossils was embedded in the heat shield of a space capsule of Foton spacecraft to test survival of the rock and the microfossils during their entry into the Earth's atmosphere (the STONE 6 experiment). This silicified volcanic sediment from the Kitty's Gap Chert (Pilbara, Australia) represented an analogue for Martian volcanic sediments of Noachian age. Additional objective of these simulation modeling experiments was to test the survival of living microorganisms (Chroococcidiopsis) within the specimen to gain an understanding whether endolithic life forms could survive atmospheric entry [Foucher *et al.* 2009]. During re-entry of Foton spacecraft, shock heating produced mineralogical alteration of the specimen with the formation of a fusion crust, cracks in the chert caused by changes of alpha quartz to beta quartz, increase in the fluid inclusion sizes, and dehydration of the hydromuscovite-replaced volcanic protoliths. The carbonaceous microfossils embedded in the chert matrix survived in the interior core of the rock away from the fusion crust. The living microorganisms were totally lost as the rock shield was only 2 cm thick whereas calculations require at least 5 cm of rock for protection of organisms to survive intense heat of entry [Foucher *et al.* 2009]. Putative large meter-sized meteoroids composed of sedimentary rocks potentially provide enough shielding to protect microscopic life forms naturally embedded in the rock interior.

**Candidates for sedimentary-origin meteorites in the literature**

Available literature contains a few reports on candidates for sedimentary-origin meteorites (hereinafter referred to as "sedimentary meteorites") with observed falls. Cross F. C. reported on three rocks found in 1947 including two grayish fine-grained sandstone specimens found in the United States that, in his opinion, deserved consideration to be of cosmic origin [Cross 1947]. Dr. Assar Hadding, a Director of the Geological Institute in Lund (Sweden), described two specimens, one of limestone and one of sandstone, that he believed were meteorites [Hadding 1940].

An unusual quartz pebble shower occurred during snowstorm in Trélex, Switzerland on February 20, 1907 [Rollier 1907]. Pebble sizes ranged from pea to hazelnut. These meteor pebbles were composed of milky quartz and were not included in large hail. The origin of pebbles was not identified with certainty. It remains unclear whether these pebbles were meteorites or they could come from the Mediterranean Sea (Islands of Hyères) or the Meseta (Spain) which is even farther away from Trélex.

An exotic gravelite sandstone boulder (Fig. 1, 2), found in association with a fresh hypervelocity disruption in the permafrost on the Stoykovich Mountain in the epicenter area of the 1908 Tunguska catastrophe, was also proposed as a candidate for sedimentary meteorite [Anfinogenov *et al*. 2014].

**Exotic rock from the epicenter of the 1908 Tunguska catastrophe**

A catastrophic collision of cosmic body with the Earth occurred above the Tunguska region of Siberia on June 30, 1908. About 2000 km$^2$ of taiga forest was devastated by shock waves and fire following the major airburst in the atmosphere. Studies propose cometary [Gladysheva 2011, 2013] and asteroid [Sekanina 2008, Chyba *et al*. 1993] origin of the Tunguska projectile. Kvasnytsya *et al*. [2013] believe it was an iron meteorite though no sizable fragments of iron meteorite have been recovered throughout the region of the catastrophe. A possible impact crater filled up by a 300 m diameter lake (Lake Cheko) has been reported ca. 8 km NNW of the Tunguska event epicenter [Gasperini *et al.* 2007, 2008, 2009] though we agree with Collins *et al.* [2008] that the lake itself is not an impact crater. Field studies suggest that numerous solutional caves (karst caves) are present throughout the region of the Tunguska catastrophe. We hypothesize that a piece of the Tunguska projectile could impact and penetrate the "roof" of such a karst cave which caused a formation of Lake Cheko in its current form. It may explain an unusual depth of the lake and the eyewitness reports stating the water bursting from under the ground after the catastrophe. Numerous alternative exotic hypotheses have been proposed to explain the nature of the 1908 Tunguska projectile that exploded in the atmosphere. However, the airburst of Chelyabinsk meteoroid [Artemieva *et al.* 2016, Popova *et al.* 2013] clearly demonstrated that the explosion in the atmosphere is a typical fate of large over meter-sized meteoroids entering the Earth's atmosphere.

A hypothesis for the existence of sedimentary meteorites was proposed in 1972 after discovery of a fresh impact-like structure near the epicenter of the 1908 Tunguska explosion. This disruption in the permafrost contained the exotic sedimentary boulder (known as John's stone or John's rock) (Fig. 1, 2); some of its splinters display a glassy surface reminiscent of fusion crust [Anfinogenov 1973, Anfinogenov *et al.* 1998a, 1998b, 2014].

John's rock is located on the Stoykovich Mountain near the epicenter of the 1908 Tunguska catastrophe. Research excavations and modeling studies suggest geologically recent hypervelocity fall of this rock. The pattern of permafrost destruction suggests hypervelocity collision and lateral ricochet of John's rock from the dense geological deposits with further deceleration and breakage producing the ~50-$m^3$ impact groove in the permafrost (Fig. 1 C and D). Landing velocity of John's rock is estimated to be at least 547 *m*/s [Anfinogenov *et al*. 2014] and may be much higher. The fall occurred in poorly consolidated ground [Anfinogenov *et al*. 2014] which presents challenges due to the lack of commonly accepted markers of hypervelocity impact in such targets [French and Koeberl 2010]. The unique phenomenon of John's rock may extend phenomenology of impact structures and contribute to identification of markers for small- and medium-scale impacts in unconsolidated or poorly consolidates targets.

John's rock is composed of abyssal highly silicified gravelite sandstone (~99% $SiO_2$). Outer surface of some splinters of this rock shows continuous glassy coating similar to fusion crust. However, most part of glassy coating and fusion crust was stripped due to the intense rolling of the rock through the ground. There is a clear consistency in geometry of meteoroid flight trajectory, locations of the John's rock fragments and cleaved pebbles, and directions of impact groves left by its largest fragments decelerating in the permafrost [Anfinogenova 2017a]. John's rock locates on quaternary deposits at the top of the Stoykovich Mountain and is mineralogically exotic to the territory within at least hundreds of kilometers around the epicenter [Sapronov 1986]. This rock significantly differs from local tufogenic sandstones, the fact which has been discussed in detail before [Anfinogenov *et al.* 2014, Sapronov 1986]. There are no signs of past glaciation throughout the region of the 1908 Tunguska catastrophe [Sapronov 1986] so John's rock could not be brought to the top of the Stoykovich mountain by any glacier. Field studies and decoding of the aerial survey photographs covering area within 40 km from the epicenter shows the absence of active diatremes [Anfinogenov *et al.* 2014] so this rock could not be ejected from a diatreme.

John's rock did not produce a typical impact crater. Instead it formed the massive pipe and the groove in the water-rich permafrost (Fig. 1 C and D). The absence of classic impact crater [Anfinogenova 2017a] is similar to several other cases when large meteorites fall on hydric soils, sands, or mountain shoulders sloping down along the direction of motion of cosmic body producing atypical impact formations. For example, among thousands of fragments of the 1947 Sikhote-Alin meteorite, several large fragments fell on frozen hydric soils forming up to 8-m long narrow canals with merely small entry holes (Fig. 3) [Krinov & Fonton, 1959].

Similarly, the largest fragment of the Chelyabinsk meteorite, a ~570 kg rock, did not form a significant impact crater at 20 m deep bottom of Lake Chebarkul despite high impact velocity for this fragment was estimated to be 19 km/s [Chapman 2013, Popova 2013]. We find this value of velocity overestimated as it corresponds to kinetic energy of 25 tons of TNT which would be significantly more destructive. Perhaps, both the largest Chebarkul fragment of the Chelyabinsk meteorite and John's rock as a putative fragment of Tunguska landed at low supersonic speeds without producing major impact craters. The failure to find well-defined impact structures associated with the 1908 Tunguska event may be also partially explained by the fact that the airburst of the meteoroid caused major earthquake up to about magnitude 8 which could largely level the disruptions produced in poorly consolidated local ground by the fallen fragments of Tunguska [Anfinogenov and Budaeva, 1998].

Detailed study of the structure, mineralogy, and chemistry of John's rock [Bonatti *et al.* 2015] confirms that the rock originated by silica deposition from hydrothermal solutions as was obvious from the very beginning after its discovery. Recent oxygen isotope data suggest that the precipitation of $SiO_2$ could have occurred in equilibrium with hydrothermal water ($\delta^{18}O_w \approx -16$ ‰) at the temperature of about 80°C [Bonatti *et al.* 2015]. High precision triple oxygen isotope data reveal that this rock is inconsistent with the composition of known Martian meteorites [Haack *et al.* 2015, Bonatti *et al.* 2015]. However, it is notable that this Tunguska boulder, in addition to $SiO_2$, contains traces of a Ti-oxide phase and its bulk composition [Bonatti *et al.* 2015] is similar to the composition inferred from APXS data [Squyres *et al.* 2008] for the silica deposits of the Gusev crater on Mars. A paper by Bonatti *et al.* [Bonatti *et al.* 2015] contains a section with detailed review of sedimentary rocks and hydrothermal activity on Mars.

Data of the field studies, research excavations, and modeling suggest hypervelocity fall of John's rock. Unlike these, instrumental data cannot unambiguously favor terrestrial or meteoric origin of John's rock. There are pros and contras [Bonatti *et al.* 2015] of extraterrestrial origin of this boulder. However, the presence of the fusion crust-like surface on the fragments and the consistent signs of hypervelocity impact associated with John's rock represent direct evidence for its meteoric nature [Anfinogenov *et al.* 2014]. Further in-depth study of this rock should be undertaken including the thermoluminescence analysis, rock age determination, and the comparison of John's rock signatures with those of similar terrestrial and extraterrestrial rocks. The site of impact requires comprehensive interdisciplinary field examination.

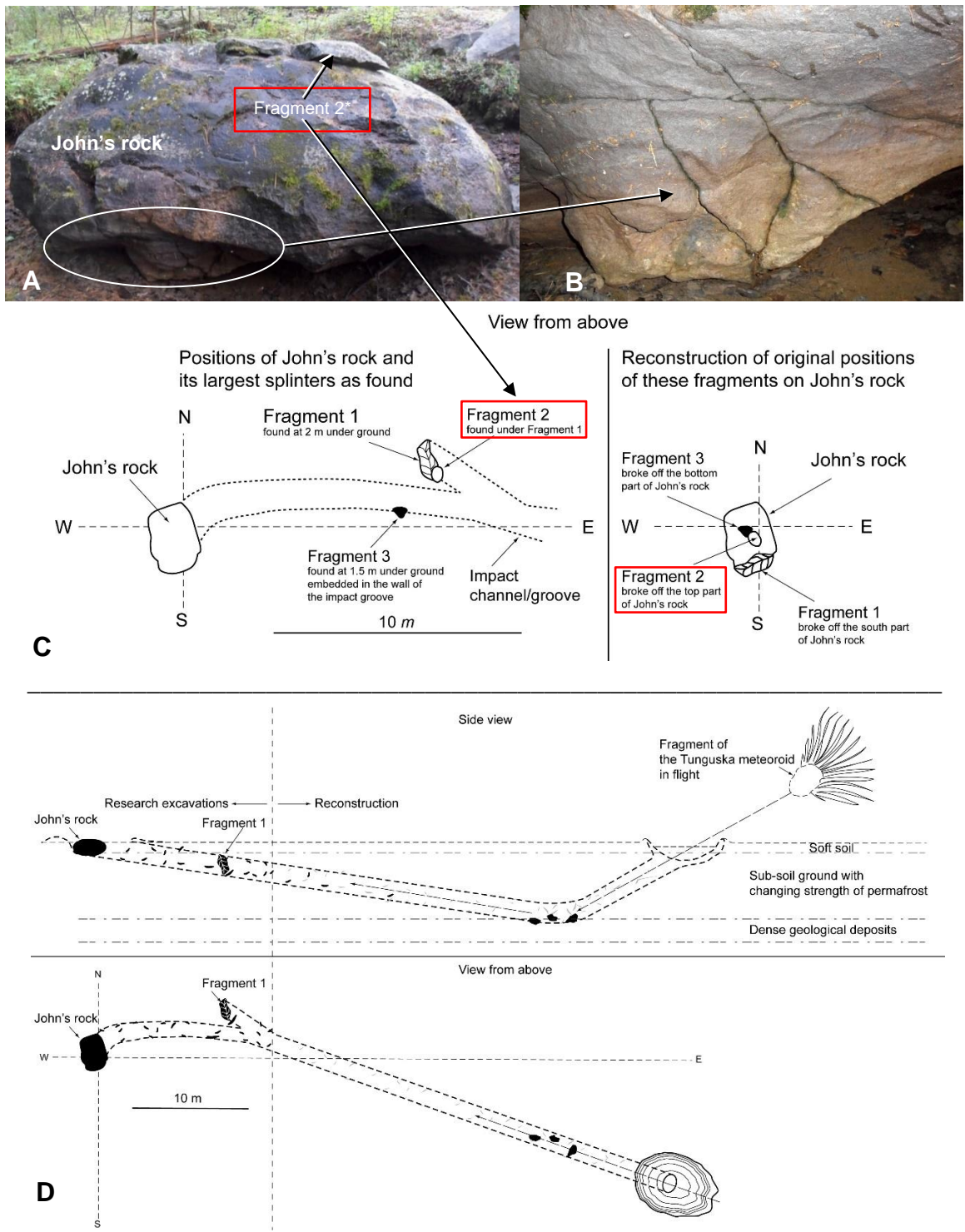

Fig. 1. *A: Photo of John's rock (2015). B: Signs of melting; possible shatter cones. C: Scheme of John's rock and its largest fragments as they were found at the Stoykovich Mountain in 1972 and their reconstruction as before breakage. D: Scheme of side and top views of hypervelocity entry, ricochet, and breakage of John's rock. Please see ref. [Anfinogenov et al. 2014] to find detailed description and images of Fragment 1.*

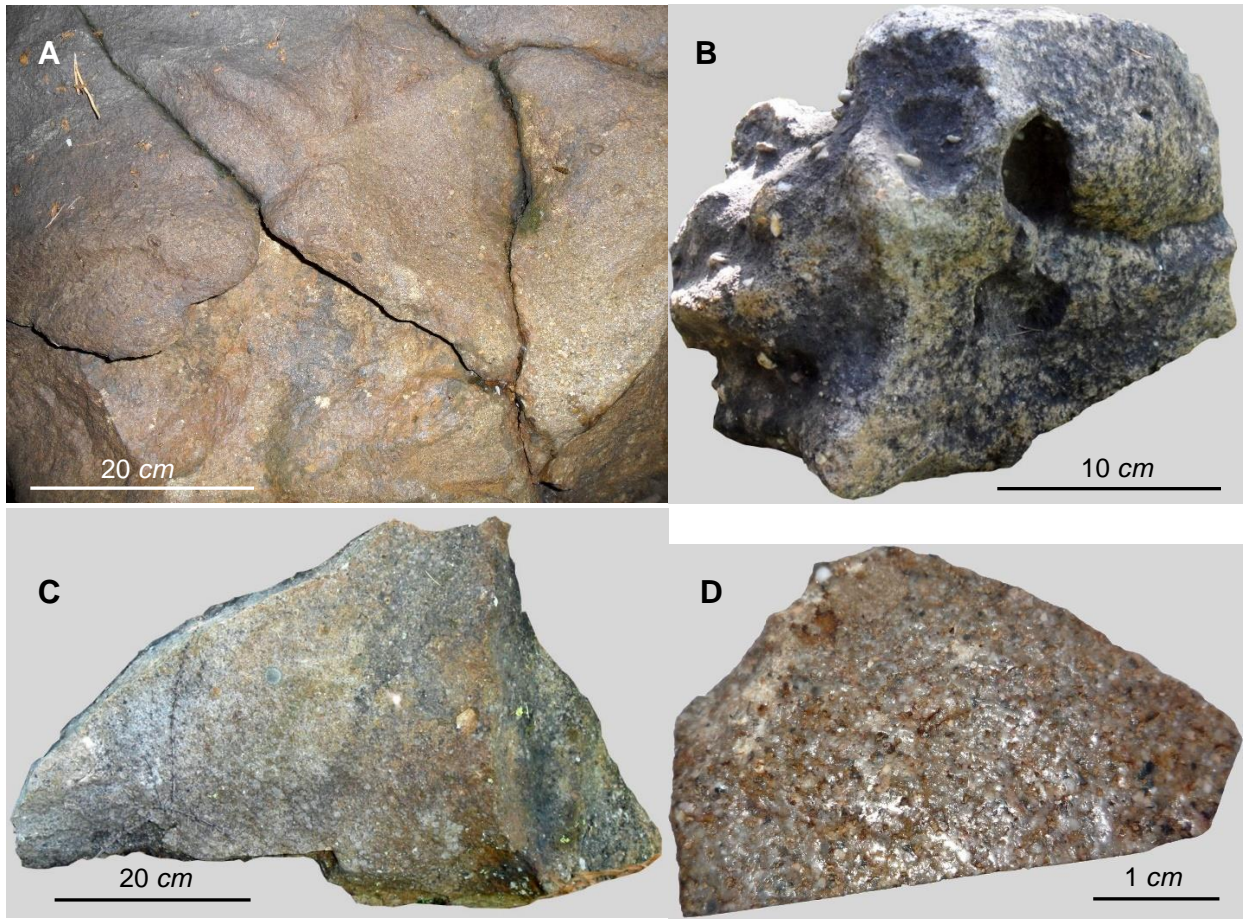

Fig. 2. *Fragments of John's rock bearing signs of hypervelocity fall and impact. A: Signs of fusion along the fracture lines and shatter cone-like structures at the bottom part of John' rock. B: Regmaglypts on the surface of ~10-kg fragment. C: Shear-fractured fragment (~50 kg) broken off the main boulder due to the fall. D: Glassy cover on one of the specimens of John's rock. Photos are made in 2014 and 2015.*

John's rock is the first macroscopic candidate for a fragment of the Tunguska projectile and for a new type sedimentary meteorite composed of silica-rich metamorphic rock. The phenomenon of John's rock complies with some criteria of impact structures though the commonly accepted markers of small- and medium-scale impacts in unconsolidated or poorly consolidated targets remain to be defined [French and Koeberl 2010]. There are melted, shocked, and brecciated rocks associated with John's rock; some of them fill the hypervelocity disruption in permafrost, and others are transported to some distances from the source impact funnel [Anfinogenov et al. 2014]. Shock deformations are present in macroscopic form [Anfinogenov et al. 2014] and in microscopic forms (e.g., deformation lamellae in quartz) [Bonatti et al. 2015]. Overall macroscopic data and mathematical calculations show that John's rock bears signs of a hypervelocity fall whose trajectory is fully consistent with the trajectory calculated from the forest fall data and observations of the Tunguska meteoroid flight [Anfinogenov et al. 2014].

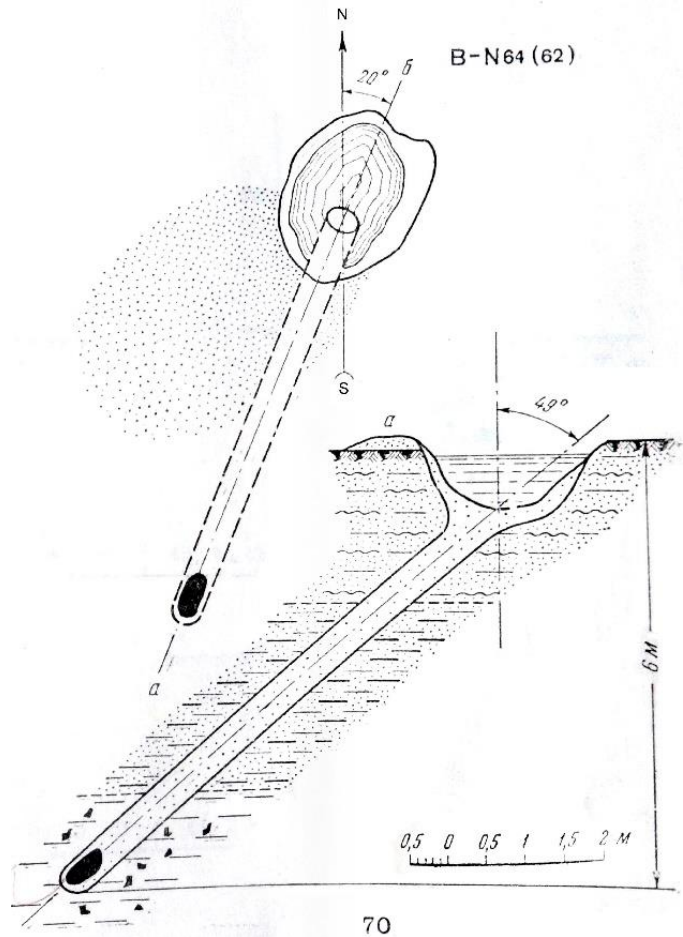

Fig. 3. *An example of the absence of typical impact crater when a fragment of the Sikhote-Alin meteoroid (1947) fell on wet sandy ground. Only a small funnel (size of 2.3 m x 1.8 m and depth of 0.9 m) was produced by this fragment (255.6 kg) of the Sikhote-Alin iron meteorite fallen in 1947. The fragment fell on wet ground composed of soil (0.2 m), clay (1.8 m), and then sand mixed with red clay. An intact individual exemplar of meteorite was recovered from the bottom of the channel at depth of 6 m [Krinov & Fonton, 1959].*

Inferred extraterrestrial origin of the Tunguska boulder [2014] is compatible with the presence of hydrothermal silica-rich deposits on Mars [McLennan 2003, Bandfield *et al.* 2004, Milliken *et al.* 2008, Squyres *et al.* 2008, Smith *et al.* 2012] as well as with the presence of liquid water [Heller *et al.* 2015, Saur *et al.* 2015, Steinbrügge *et al.* 2015, Tanigawa *et al.* 2014, Vance *et al.* 2014] and hydrothermal activity [Hsu *et al.* 2015, Postberg *et al.* 2011] on several other bodies of the Solar System such as icy moons of Jupiter and Saturn. Mars is not the only candidate parent body for putative sedimentary meteorites. We hypothesize that sedimentary rocks may form in the presence of water flows generated by tidal and volcanic forces in the oceans on the satellites of the giant planets.

**Hypothesis of Tunguska rubble-pile asteroid partially composed of extraterrestrial sediments**

In mid 1960s, J. Anfinogenov carried out a microscopy study of burn injuries detected in the 1908-growth rings of the branches from the larch trees that survived the catastrophe in the epicentral area. He observed the presence of microspherules and visually heterogeneous fused microparticles. Similar particles were found in the peat layer of 1908. In particular, the discovery of a significantly higher content of glassy silicate microspherules precipitated from the atmosphere was reported [Dolgov et al. 1973]. Compared with the adjacent peat layers, the peat layer of 1908 contains up to hundredfold-higher count of gray and colorless transparent silicate microspherules in hundreds of samples taken over the entire area of the Tunguska catastrophe. Data of neutron activation analysis show that chemical composition of these microspherules is distinct from that of industrial glass, local terrestrial microparticles, known stony meteorites, tektites, and Moon rocks [Kolesnikov et al. 1976]. The quartz grains were also found in sediment cores collected from Lake Cheko [Gasperini et al. 2009] in the epicentral area suggesting that they might have resulted from dust produced by the explosion in the atmosphere of the main body if the Tunguska cosmic body were silica-rich. The presence of this silica anomaly is consistent with the hypothesis on meteoritic origin of silica-rich John's rock.

However, the presence of PGE anomaly was also reported [Rasmussen et al. 1999, Hou et al. 2004] for Tunguska though these reports were based on data from an insignificant number of samples. Peat cores from singular topographic locations were tested in these studies. Therefore, two distinct anomalies have been reported in the 1908 peat layer over the area of interest: (i) above-mentioned anomalous abundance of glassy silicate microspherules precipitated from the atmosphere and (ii) PGE anomaly.

Two explanations may be proposed to reconcile these findings. According to the first explanation, two impact events involving silica-rich impactor and chondritic or cometary projectile happened independently within short period of time. The second explanation proposes the rubble-pile asteroid hypothesis [Anfinogenova et al. 2017b] and is more plausible in our opinion. It suggests that the impactor had a complex conglomerate composition where multiple pieces merged due to either collision of parent asteroids in outer space or due to a co-ejection of different adjacent rocks from a parent cosmic body after a high-energy impact. It may be a co-ejection of enclosing bedrocks, intrusive igneous rocks, and impactor material where any of these components could partially melt into each other due to mega-impact on parental planetary body (Fig. 4). Such processes could produce a rubble-pile asteroid whose fall caused dual anomaly at the 1908 impact site in Tunguska.

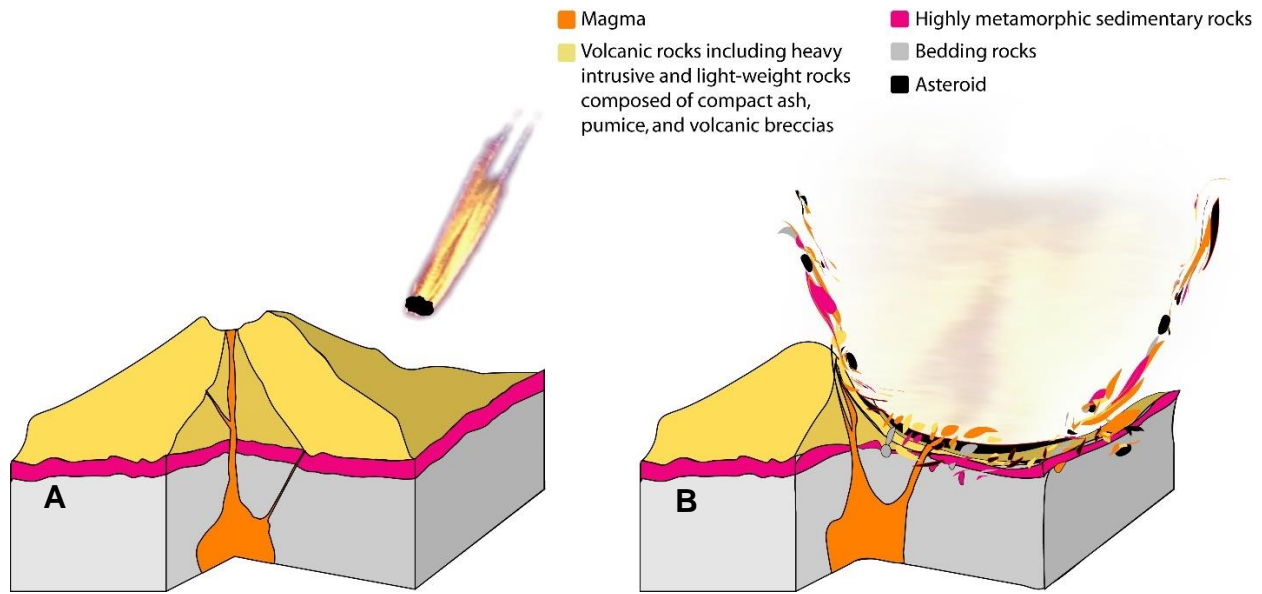

Fig. 4. *A: Bombardment of the volcano by an asteroid. B: Ejection of complex-composition pro-asteroid material into space due to bombardment.*

Interestingly, no macroscopic pieces of chondritic or cometary projectile have ever been found in the area of the 1908 Tunguska catastrophe. Chyba C. F. *et al*. [1993] reported that carbonaceous asteroids and especially comets are unlikely candidates for the Tunguska object and that the Tunguska event represents a typical fate for stony asteroids tens of meters in radius entering the Earth's atmosphere at common hypersonic velocities [Chyba *et al*. 1993]. In this regard, John's rock may be considered a sound macroscopic candidate for a stony impactor though of a previously unknown type. This hypothesis is consistent with John's rock phenomenon bearing the numerous signs of hypervelocity impact and glassy fusion crust-like surface on some splinters [Anfinogenov and Budaeva 1998, Anfinogenov et al. 2014]. It is also consistent with the discovery of the quartz grains in sediment cores collected from Lake Cheko [Gasperini et al. 2009] in the epicentral area and the glassy silicate microspherules anomaly associated with the entire area of the 1908 Tunguska catastrophe [Dolgov et al 1973, Kolesnikov et al. 1976].

Data showing the presence of two distinct anomalies suggest that the 1908 Tunguska cosmic body may be a rubble pile asteroid partially composed of a material representing planetary crust of Earth-like planet.

Conglomerate nature of the Tunguska meteoroid may explain the features of its atmospheric disintegration. Starting from 1965, integrative Tunguska field studies, modeling, and expert examination of the aerial photography (1938, 1949 air surveys) including aerial-image-3D-visualization (stereo pairs) produced data showing that the airburst was not single-point; the initial blast wave had a shape of a spindle or the tip of a spear [Anfinogenov 1966, Anfinogenov and Budaeva 1998]. According to these data, Tunguska meteoroid probably disintegrated into ca. 20 fragments each of which were comparable with the 2013 Chelyabinsk meteoroid. There are signs that the Tunguska cosmic body was irregular (heterogeneous) in its composition and mechanical and thermal strengths. Initial breakage of Tunguska meteoroid into the fragments resulted in the formation of the swarm of meteoroids that exploded at different altitudes ranging from 25 to 5 km above the ground. Cumulatively, these explosions produced the continuous forest fall over the area of 600 km$^2$. Small fragments (composed of slightly altered matter) remaining from those meteoroids could reach the ground similarly to the 2013 Chelyabinsk. According to data of the aerial photography (1938, 1949) covering the area of the Tunguska catastrophe, the regenerating holes were present in the Southern Bog (Yuzhnoe Boloto) and near the epicenter at time of air surveys [Anfinogenova et al. 2016].

**Potential parent bodies**

Sandstones can only form on a parent body with liquid water. In the Solar System, many bodies without a significant atmosphere have abundant liquid water [Heller *et al.* 2015, Saur *et al.* 2015, Steinbrügge *et al.* 2015, Tanigawa *et al*. 2014, Vance *et al.* 2014, Greenwood et al. 2005]. Hydrothermal processes occur, in particular, on Enceladus [Hsu *et al.* 2015, Postberg *et al*. 2011]. The presence of powerful tidal currents of water in subsurface oceans on icy satellites of Saturn and Jupiter may provide conditions necessary and sufficient for formation of sedimentary and metamorphic rocks including sandstones and pebble conglomerates with a variety of grain sizes.

The Saturnian moon Enceladus may be considered a candidate parent body for sedimentary meteorites composed of highly metamorphic rock due to (i) the presence of global hydrothermal activity, (ii) the presence of powerful tidal currents of water in subsurface oceans potentially resulting in formation of sediments, and (iii) the past history of large-scale impacts explaining ejection of Enceladus' crust fragments into space. Interestingly, the plume of Enceladus emits nanometre-sized $SiO_2$ (silica)-containing ice grains [Hsu *et al.* 2015] formed as frozen droplets from a liquid water reservoir contacting with silica-rich rock [Postberg *et al.* 2011]. Characteristics of these silica nanoparticles indicate ongoing high temperature (>90 °C) global-scale geothermal and hydrothermal reactions on Enceladus favored by large impacts [Hsu *et al.* 2015]. Enceladus has a differentiated interior consisting of a rocky core, an internal ocean and an icy mantle. Simulation studies suggest that large heterogeneity in the interior, possibly including significant core topography may be due to collisions with large differentiated impactors with radius ranging between 25 and 100 km. Impacts played the crucial role on the evolution of Enceladus [Monteux *et al.* 2016] and similar effects on evolution are very likely on the other moons of Saturn as well as on other planetary objects, such as Ceres [Davison *et al.* 2015, Ivanov 2015].

A putative subsurface water ocean is present on Ganymede. Iron core of Ganymede is surrounded by a silicate rock mantle and by a globe-encircling, briny subsurface water ocean with alternating layers between high pressure ices and salty liquid water [Saur *et al*. 2015, Steinbrügge *et al*. 2015, Vance *et al*. 2014]. If Ganymede or Callisto had acquired their $H_2O$ from newly accreted planetesimals after the Grand Track [Mosqueira *et al.* 2003], then Io and Europa would be water-rich, too [Heller *et al.* 2015, Tanigawa *et al.* 2014].

Tidal dissipation and tidal resonance in icy moons with subsurface oceans are major heat sources for these icy satellites of the giant planets [Kamata *et al.* 2015]. Tidal forces generate heat and currents of liquid water or brine powerful enough to produce sediments that undergo metamorphic transformations due to hydrothermal activity. Galilean satellites have great potential as targets for astrobiological exploration [Greeley and Morrison 2003].

There is also evidence for aeolian bedforms and progressive induration in Medusae Fossae Formation transverse aeolian ridges on Mars [Kerber & Head 2012]. There is groundwater activity [Michalski et al. 2013] and pebbles on Mars [Jerolmack 2013]. Martian fluvial conglomerates are discovered at Gale Crater [Williams et al. 2013].

Therefore, candidate parent bodies for hypothetical sedimentary meteorites comprise, but are not limited by the Earth, Mars, Enceladus, Ganymede, and Europa.

**Hypothetical planet Phaeton as possible parent body**

After discovery of the sedimentary boulder associated with hypervelocity disruption of the ground in the epicenter of the 1908 Tunguska catastrophe, it was hypothesized [Anfinogenov 1973, 2017, Anfinogenov and Budaeva 1998] that sedimentary meteorites may come to the Earth from a hypothetical planet called Phaeton [McSween 1999]. Findings of the so-called Martian meteorites and some pseudo-meteorites belonging to upper-crust rocks (volcanic and highly-metamorphic igneous and sedimentary rocks) from Mars-like planets provide rationale for reconsideration of the hypothesis on past existence of a planet between the orbits of Mars and Jupiter as a parent body of the asteroid belt. Nature of the hypothetical planet and mechanism of its total explosive disintegration remain the main questions. Various mechanisms were proposed to answer the challenges of the exploded planet hypothesis [Van Flandern 2007]. Considering interorbital structural characteristics of the Solar system, we believe that two or even three planets could be spaced between the orbits of Mars and Jupiter. Providing certain geometry of their orbits and influence of the giant planet Jupiter, these neighbor planets could close with each other up to catastrophic collision.

Based on mathematical modeling with iterative refining, we propose a scenario [Anfinogenov et al. 2017] involving two hypothetical planets Phaeton I and Phaeton II (Fig. 5) with the following characteristics: (i) masses comparable with the mass of Mars; (ii) average distances of 2.4 A.U. between the Sun and Phaeton I and 3.95 A.U. between the Sun and Phaeton II; (iii) elliptic orbits like those of Pluto and Mercury with the major axes in the ecliptic plane; (iv) orbital plane inclinations of 15° relative to the ecliptic plane, and the angle of 30° between the orbital planes of these planets; (v) similar Phaeton II's perihelion and Phaeton I's aphelion distances (2.9 ± 0.1 A.U. from these planets to the Sun, respectively) and the distance between these planets of ≤0.1 A.U. at their closest approach to each other.

The catastrophic collision was possible in case of spatial and temporal co-occurrence of Phaeton II's perihelion and Phaeton I's aphelion if the courses of these planets were intersecting at 30° angle; if individual orbital velocities of Phaeton I and Phaeton II were ~16 and ~20 km/s, respectively; and if their closing velocity was ~9 km/s at the moment of the collision. If this was the case, different variants of a spacial arrangement of Phaeton I and Phaeton II physical bodies relative to each other were possible including a scenario where Phaeton I collided with a rear hemisphere of Phaeton II (billiard ball impact type). If closing velocity was ~9 km/s, the impulse of impact force and the kinetic energy were sufficient for disintegration of both planets and dispersion of their fragments with the velocities exceeding second cosmic speed for their masses. In such a case, a significant portion of the mass from Phaeton II and its fragments might acquire an additional orbital acceleration accounting for the stretching of their orbit up to the Jupiter orbit with a possible capture of some Phaeton II fragments by Jupiter so these fragments assumed orbital motion around Jupiter and even the entry into the Jovian atmosphere. A significant portion of the mass from Phaeton I might undergo deceleration which caused the exit of its fragments from orbit and the acceleration towards the Sun and the Earth-type planets with a possible fall on them. Middle part of the zone where two planets collided could form the asteroid belt.

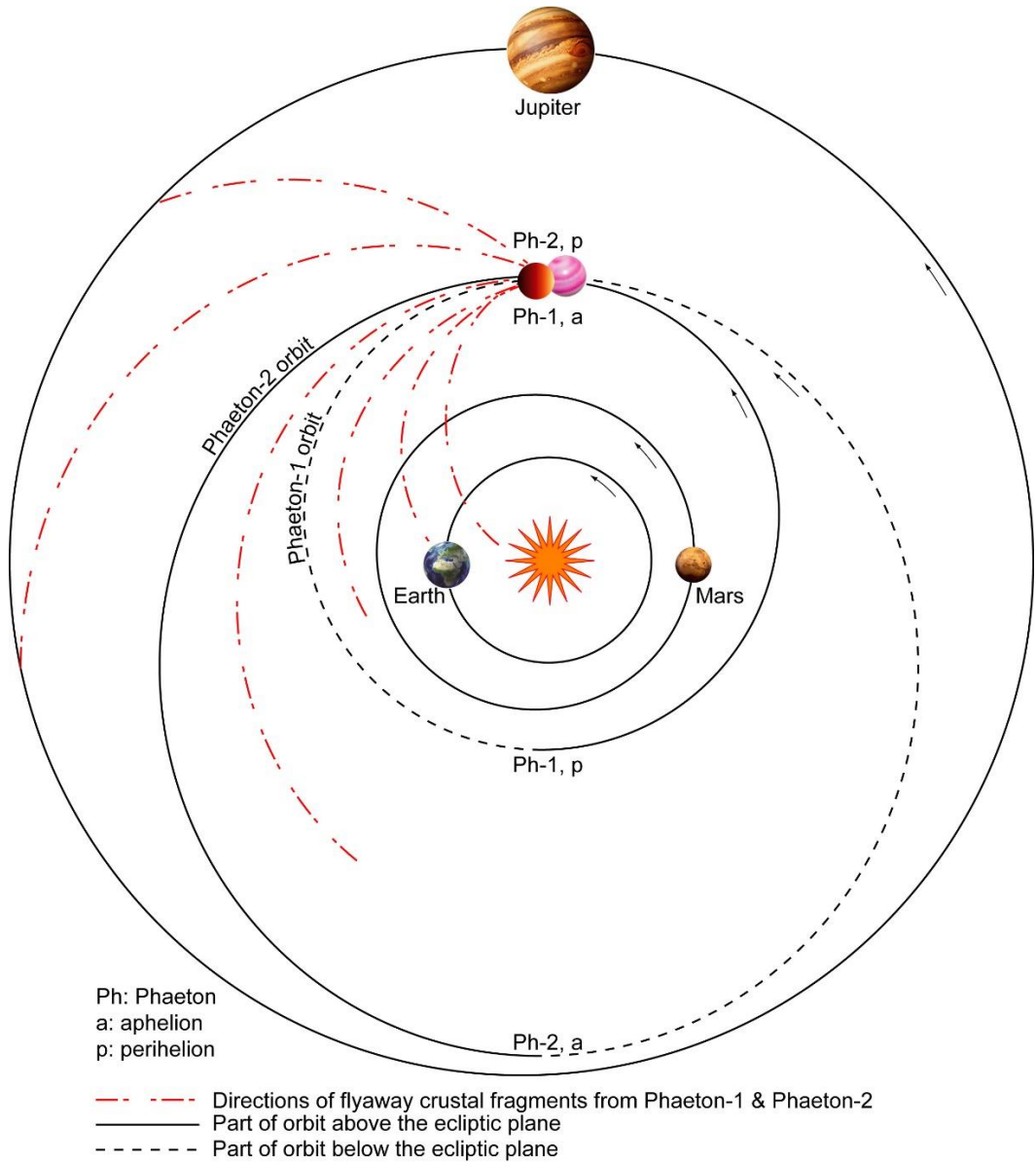

Fig. 5. S*cenario of the moment of collision of hypothetical planets Phaeton I and Phaeton II [Anfinogenov et al. 2017]*.

We believe that the layered structure of Mars' moon Phobos may suggest its origination from the planetary crust of Phaeton I. This notion agrees well with work of Simioni E. *et al.* [2015] who proposes the explanation of the observed distribution of the grooves on Phobos as remnant features of an ancient parent body from which Phobos could have originated after a catastrophic impact event [Simioni E. *et al.* 2015]. We hypothesize that Olympus Mons, the largest volcano in the Solar system, could form when the other large fragment of Phaeton I collided with Mars and produced gigantic impact hole in its planetary crust.

Proposed reconstruction of the initial planetary structure of the Solar system and its partial disruption are compatible with the structural and dynamic characteristics of the preserved parts of the Solar system. Identification of progenitors for the meteorites and the bodies from the asteroid belt as well as determination of cosmic age of their formation require to consider that some of them may belong to the rocks originating from planetary crusts of hypothetical planets Phaeton I and Phaeton II.

Therefore, Hypothetical planets Phaeton I and Phaeton II that possibly existed in the past may represent parental bodies of putative new-type planetary meteorites including Martian-like meteorites and meteorites belonging to upper-crust rocks such as volcanic and highly-metamorphic igneous and sedimentary rocks.

**Isotope tests of candidate extraterrestrial sedimentary rocks**

Isotopic characterizations of the exotic boulder from the epicenter of the 1908 Tunguska event are essential and further elemental and isotopic characterizations of this rock are highly encouraged. High precision triple oxygen isotope data reveal that this rock is inconsistent with the composition of known Martian meteorites [Haack *et al*. 2015, Bonatti *et al*. 2015]. However, numerical results of isotopic characterizations, though important on their own, cannot prove or disprove the origin of the rock. Indeed, no rock samples have ever been delivered from Mars or from other cosmic bodies with abundant water to the Earth before. Isotopic compositions of Martian sedimentary rocks have never been tested by Mars rovers and remain unknown. All which was tested up to day were the Martian meteorites. Meanwhile, diversities in the rock-forming processes and in the corresponding rock types within planets are insufficiently studied. A significant heterogeneity in $\Delta^{17}O$ in rocks of different types has been reported [Wang *et al*. 2013]. Pack and Herwartz [Pack *et al*. 2014, 2015] provide evidence that the concept of a single terrestrial mass fractionation may be invalid on small-scale. They conclude that mineral assemblages in rocks fall on individual "rock" mass fractionation lines with individual slopes and intercepts.

We believe the same may be true for other bodies of the Solar System such as large moons of Jupiter and Saturn, and large asteroids. An idea of separate long-lived silicate reservoirs on Mars is supported by radiogenic isotope studies [Borg *et al*. 1997, 2003]. The distinct $\Delta^{17}O$ and $\delta^{18}O$ values of the silicate fraction of NWA 7034 compared to other SNC meteorites support the idea of distinct lithospheric reservoirs on Mars that have remained unmixed throughout Martian history [Agee *et al.* 2013, Ziegler *et al.* 2013]. Isotopic heterogeneity, including that in the noble gases, can be significant in the Martian mantle. Models for accretion and early differentiation of Mars were tested with chronometers several of which provided evidence of very early isotopic heterogeneity preserved within Mars [Halliday *et al.* 2001]. If significant heterogeneities are reported for known Martian meteorites which are all igneous rocks, then even greater isotopic heterogeneities may exist for the rocks of different types within the planet.

Terrestrial loss or contamination in the noble gas signatures should be also considered for any meteorite including Martian meteorites that spent time in terrestrial environment, especially in extreme weather conditions. In our opinion, macroscopic evidence from field studies suggesting hypervelocity fall of the rock should be taken into careful consideration. The findings supporting hypervelocity fall of John's rock in Tunguska cannot be denied based on studying its isotopic signatures. Considering the lack of knowledge on isotopic signatures of several admissible cosmic bodies that may be the places of origin for putative sedimentary meteorites, data suggesting hypervelocity fall of the rock must be taken into account.

**Summary**

We conclude that the list of candidate parent bodies for putative sedimentary meteorites includes, but is not limited by the Earth, Mars, Enceladus, Ganymede, and Europa. John's rock from the 1908 Tunguska catastrophe epicenter is a candidate sedimentary meteorite composed of silica-rich sedimentary rock available for examination. The macroscopic signs of geologically fresh hypervelocity pipe-like impact associated with John's rock, the presence of microscopic deformation lamellae, fusion crust-like glassy surface, location of this boulder near the epicenter of the 1908 Tunguska catastrophe, and its mineralogy alien to the region represent extraordinary evidence for meteoric origin of this rock. Tunguska may be a case for an identification of the markers for small- and medium-scale impacts in poorly consolidated or unconsolidated ground. Planetary science and astrobiology will significantly benefit from identification of planetary-origin meteorites with non-igneous composition.